\documentstyle[12pt]{article}
\begin{document}
\pagestyle{empty}
\vskip .7cm
\begin{center}

{\bf {Cohomological aspects of Abelian gauge theory}}

\vskip 2cm

{\bf R. P. Malik} 
\footnote{ E-mail address: malik@boson.bose.res.in }\\
{\it S. N. Bose National Centre for Basic Sciences,}\\
{\it Block-JD, Sector III, Salt Lake, Calcutta- 700 091, India}\\

\vskip .5cm

{\bf ABSTRACT}

\end{center}
 
We discuss some aspects of cohomological properties of a 
two-dimensional free Abelian gauge theory
in the framework of BRST formalism. We derive the 
conserved and nilpotent BRST- and 
co-BRST charges and express the Hodge decomposition
theorem in terms of these charges and a conserved
bosonic charge corresponding to the Laplacian
operator. It is because of the topological nature of free
$U(1)$ gauge theory that the Laplacian operator goes to zero 
when equations of motion are exploited.
We derive two sets of topological invariants 
which are related to each-other by a certain kind of 
duality transformation and express the Lagrangian density of this theory as
the sum of terms that are BRST- and co-BRST invariants. Mathematically,
this theory captures together some of the key features of Witten- and
Schwarz type of topological field theories.\\

\baselineskip=16pt

\vskip 1cm
\newpage

\pagestyle{headings}

\noindent 
{\bf 1 Introduction}\\

\noindent
One of the key theorems in the mathematical aspects of cohomology 
is the celebrated 
Hodge decomposition theorem defined on a compact manifold.
This theorem states that
any arbitrary $p$-form $f_{p}$ on this manifold can be decomposed into a
harmonic form $\omega_{p}$ $( \Delta \omega_{p} = 0, d 
\omega_{p} = 0, \delta
\omega_{p} = 0)$, an exact form 
$ d\; g_{p-1}$ and a co-exact form $ \delta\; h_{p+1}$:
$$
\begin{array}{lcl}
f_{p} = \omega_{p} + d\; g_{p-1} + \delta\; h_{p+1}
\end{array}\eqno(1.1)
$$
where $\delta (= \pm\; {*}\; d \;{*})$ is the Hodge dual of $d$ 
(with $ d^2 = 0,\delta^2 = 0 )$
and Laplacian $\Delta$ is defined as $ \Delta = ( d + \delta )^2
= d \delta + \delta d $ [1-4]. So far, the analogue of $d$ has been found
out as the local, conserved and nilpotent ($Q_{B}^2 = 0 $) 
Becchi-Rouet-Stora-Tyutin
(BRST) charge $Q_{B}$ which generates a nilpotent
BRST symmetry for a locally gauge invariant 
Lagrangian density in any arbitrary
dimension of spacetime. The physical state condition
$Q_{B} \;| phys >\; = \;0 $ leads to the annihilation
of physical states in the quantum Hilbert space by the 
first-class constraints
of the original gauge theory. This requirement is essential for the consistent
quantization of a theory endowed with the first-class 
constraints (see, {\it e.g.}, [5-10])
\footnote{ Attempts have also been made to discuss the
second-class constraints in the framework of  BRST formalism 
(see, {\it e.g.},
[11,12] and references therein).}. It will be an interesting idea to explore
the possibility of finding out the {\it local} conserved charges 
corresponding to
$\delta$ and $\Delta$ so that a complete 
physical understanding of BRST cohomology
and Hodge decomposition can emerge in the quantum Hilbert space of states.

The purpose of the present work is to provide some physical 
interpretations to the analogues of
$\delta$ and $\Delta$ in the language of nilpotent (for $\delta$), local, 
covariant and continuous symmetry properties of a free $U(1)$ gauge 
theory described by the BRST invariant Lagrangian densities and show that
this theory is a tractable field theoretical model for the Hodge theory
in two $ (1 + 1) $ 
dimensions of spacetime. Some very interesting and illuminating
attempts [13-16] have been made towards this 
goal for the Abelian as well as non-Abelian 
gauge theories in any arbitrary dimension of spacetime.
However, the symmetry transformations turn out to be nonlocal
and noncovariant. In the relativistic covariant formulation, the symmetry
transformations turn out to be even non-nilpotent and they
become nilpotent only when some restrictions are imposed [17]. 
We shall demonstrate that for the two
dimensional  BRST invariant free $U(1)$ gauge theory, a conserved and nilpotent 
co(dual)-BRST charge $Q_{D}$ ($i.e.$, the analogue of $\delta$)
can be defined which corresponds to a new
local, covariant, continuous and nilpotent
symmetry transformation under which  the gauge-fixing term 
$ \delta A = (\partial \cdot A)$ 
\footnote { Here one-form $ A =
A_{\mu}\; dx^{\mu}$ defines the vector potential $A_{\mu}$ of the $U(1)$
gauge theory. Furthermore, it can be easily seen that 
the gauge-fixing term $ (\partial \cdot A)= \delta\; A$
is the Hodge dual of the two-form $ F = d A $ in the Abelian 
$U(1)$ gauge theory in any arbitrary dimension of spacetime
(see, {\it e.g.}, Ref. [2]).} remains invariant.
This should be compared and contrasted with the usual BRST transformation 
under which the two-form $ F = d A $ remains invariant in the $U(1)$ 
gauge theory. Further, we show that the anticommutator of both these charges
$ W = \{ Q_{B}, Q_{D} \}$ is the analogue of the Laplacian operator
$\Delta$ and it turns out to be the Casimir operator for the extended
BRST algebra. We implement the Hodge decomposition theorem with these 
charges and show that the requirement of the
annihilation of physical (harmonic) states by $Q_{B}$
and $Q_{D}$ is sufficient to gauge away both the degrees of freedom of a
single photon in 2D. The ensuing theory becomes topological in 
nature (as there are no propagating degrees of freedom left in the theory) [18].
In the framework of BRST cohomology and Hodge decomposition theorem,
this fact is encoded in rendering the Casimir operator $W$ to go to zero
($ W \rightarrow 0$) when equations
of motion are exploited and all the fields are assumed to fall off rapidly
at $ x \rightarrow \pm \infty $. On the contrary, for the 2D interacting
$U(1)$ gauge theory, it has been shown that $W$ does not go to zero
on the on-shell because of the presence of matter degrees of freedom in
the theory [19].
For the topological 2D free $U(1)$ gauge theory,
we derive  two sets of topological invariants 
with respect to both the conserved 
and nilpotent charges $Q_{B}$ and $Q_{D}$. These 
invariants turn out to be connected with each-other by a 
certain specific type of duality transformation.

The outline of our paper is as follows. In Sec. 2, we set
up the notations and sketch briefly the essentials of 
BRST formalism for $U(1)$ gauge theory in any arbitrary dimension of 
spacetime. Sec. 3 is devoted to the derivation of the nilpotent and
conserved (anti)dual BRST charge and the Laplacian operator in 
two dimensions of spacetime. This
is followed, in Sec. 4, by the discussion of an extended BRST algebra 
which is constituted
by six conserved charges. We discuss Hodge decomposition theorem and obtain
two sets of topological invariants in Sec. 5. 
Finally, we make some concluding
remarks  in Sec. 6.\\

\noindent             
{\bf 2 Preliminary: BRST invariant Lagrangians}\\

\noindent
We begin with the BRST invariant Lagrangian density (${\cal L}_{b}$) for the
$U(1)$ gauge theory in the Feynman gauge (see, {\it e.g.}, [5-9])
$$
\begin{array}{lcl}
{\cal L}_{b} = - \frac{1}{4} F^{\mu\nu} F_{\mu\nu}
- \frac{1}{2} (\partial \cdot A)^2 - i \partial_{\mu}\bar C \partial^{\mu} C
\end{array} \eqno(2.1)
$$
where the first term is the classical Maxwell Lagrangian density and second
and third terms are the gauge-fixing and Faddeev-Popov ghost terms 
respectively. Here the $U(1)$ gauge connection $A_{\mu}$ is defined through
the one-form $ A = A_{\mu} \;dx^{\mu} $ and the curvature term
$F_{\mu\nu} = \partial_{\mu} A_{\nu} - \partial_{\nu} A_{\mu}$ 
($ \mu, \nu = 0, 1, 2,....D-1$) is  obtained
from the two-form $ F = d A $ in any D-dimensional flat Minkowski spacetime.
Furthermore, the gauge-fixing term
$ (\partial \cdot A) = \partial_{\mu} A^{\mu} \equiv \delta A$, is the Hodge
dual of the two-form $ F = d A $
and $ \bar C ( C) $ are the anti
(ghost) fields. The following on-shell ($ \Box C = 0 $)
nilpotent $(\delta_{b}^2 = 0)$ symmetry
transformations 
$$
\begin{array}{lcl}
\delta_{b} A_{\mu} &=& \eta \partial_{\mu} C, \qquad \delta_{b} C = 0,
\quad \delta_{b} F_{\mu\nu} = 0 \nonumber\\
\delta_{b} \bar C &=& - i \eta (\partial \cdot A), \qquad
\delta_{b} (\partial \cdot A) = \eta \Box C
\end{array} \eqno(2.2)
$$
lead to the derivation of a conserved and 
on-shell nilpotent BRST charge $Q_{b}$
$$
\begin{array}{lcl}
Q_{b} = 
 {\displaystyle \int d^{D-1} x}\; 
\bigl [\; \partial_{0} (\partial \cdot A) C - 
(\partial \cdot A) \partial_{0} C\; \bigr ]
\end{array} \eqno(2.3)
$$
where $\eta$ is an anticommuting ($ \eta \;C = - C\; \eta, 
\eta\; \bar C = - \bar C\; \eta$)
spacetime independent infinitesimal parameter.
Introduction of an auxiliary field $B$ in the Lagrangian density (2.1)
$$
\begin{array}{lcl}
{\cal L}_{B} = - \frac{1}{4} F^{\mu\nu} F_{\mu\nu} + B (\partial \cdot  A)
+ \frac{1}{2} B^2 - i \partial_{\mu} \bar C \;\partial^{\mu} \;C
\end{array} \eqno(2.4)
$$
enables the validity of an off-shell nilpotent $(\delta_{B}^2 = 0)$
symmetry transformations
$$
\begin{array}{lcl}
\delta_{B} A_{\mu} &=& \eta \;\partial_{\mu} \;C,\;\; \qquad
\;\;\;\delta_{B} F_{\mu\nu} = 0, \;\;\qquad \;\;\; 
\delta_{B} C = 0 \nonumber\\
\delta_{B} \bar C &=& i \;\eta\; B, \qquad \delta_{B} B = 0,
\qquad \delta_{B}(\partial \cdot  A)= \eta \;\Box \;C  
\end{array} \eqno(2.5)
$$
which lead to the existence of an off-shell
nilpotent and conserved BRST charge

$$
\begin{array}{lcl}
Q_{B} = {\displaystyle \int d^{(D-1)} x}\; \bigl [ B \dot C - \dot B C \bigr]. 
\end{array} \eqno(2.6)\\
$$

The invariance of the ghost action $I_{F.P.} = - i \int d^{D}x
\partial_{\mu}  \bar C  \partial^{\mu} C $ under discrete symmetry
transformations: $ C \rightarrow \pm i \bar C, \bar C \rightarrow \pm i C$
implies the existence of a conserved and nilpotent anti-BRST charge $(Q_{AB})$
which can be derived from the expressions (2.3) and (2.6) by the substitution
$ C \rightarrow \pm i \bar C $. The continuous global symmetry invariance
of the total action under the transformations: $ C \rightarrow e^{-\lambda} C,
\bar C \rightarrow e^{\lambda} \bar C, A_{\mu} \rightarrow A_{\mu},
B \rightarrow B $, 
(where $\lambda$ is a global parameter), leads
to the derivation of the conserved ghost charge $(Q_{g})$
$$
\begin{array}{lcl}
Q_{g} = - i {\displaystyle \int d^{(D-1)} x }\;
\bigl [\; C\; \dot {\bar C} + \bar C\; \dot C \;\bigr ].
\end{array}\eqno(2.7)
$$
Together, these conserved charges obey the following algebra:
$$
\begin{array}{lcl}
&& Q_{B}^2 = \frac{1}{2}\;\{ Q_{B}, Q_{B} \} =0,\; 
Q_{AB}^2 = \frac{1}{2}\;\{ Q_{AB}, Q_{AB} \}= 0 
\nonumber\\
&& \{ Q_{B},   Q_{AB} \}\; =\; Q_{B}\; Q_{AB}\; +\; Q_{AB}\; Q_{B} 
=\; 0 \nonumber\\
&& i [  Q_{g}, Q_{B} ] = + Q_{B},\;\;\; \quad\;\;\; 
i [  Q_{g}, Q_{AB} ] = - Q_{AB} 
\end{array}\eqno(2.8)
$$
where the cononical (anti)commutators for the BRST invariant Lagrangians
are exploited for the derivation of the above algebra. This algebra is valid 
for $U(1)$ gauge theory in any arbitrary
dimensions of spacetime.
It will be noticed that the anticommutator $ \{ Q_{B}, Q_{AB} \}= 0$
implies that the combined transformations  $\delta_{B} \delta_{AB}
+ \delta_{AB} \delta_{B}$ acting on any field produce no transformation
at all. Thus, anti-BRST
charge is not the analogue of the dual(adjoint) exterior derivative
$(\delta)$ for the $U(1)$ gauge theory
\footnote{ It has been demonstrated in Ref. [20] that
the anticommutator of the cohomologically higher order BRST- and 
anti-BRST charges is not
zero and it leads to 
the definition a cohomologically higher order Laplacian operator for
the compact non-Abelian Lie algebras.}.\\

\noindent
{\bf 3 Dual-BRST symmetry in two dimensions}\\

\noindent
In addition to the symmetries: $ C \rightarrow  \pm i \bar C,\; \bar C
\rightarrow  \pm i C $, the ghost action $- i 
\int \; d^2 x\;\;\partial_{\mu} \bar C \;\partial^{\mu} \;C $ in 2D 
respects another symmetry; namely,
\footnote{ We adopt here the notations in which the 2D flat Minkowski
metric is : $\eta_{\mu\nu} = $ diag $ (+1, -1) $ and $ \Box = \eta^{\mu\nu}
\partial_{\mu} \partial_{\nu} = \partial_{0} \partial_{0} -
\partial_{1} \partial_{1}, \dot f = \partial_{0} f, F_{01} = 
\partial_{0} A_{1} - \partial_{1} A_{0} = E
= F^{10}, \varepsilon_{01} = \varepsilon^{10} = +1, 
(\partial \cdot A) = \partial_{0} A_{0} - \partial_{1} A_{1}$. }
$$
\begin{array}{lcl}
\partial_{\mu} \rightarrow  \pm\; i\; \varepsilon_{\mu\nu}\; \partial^{\nu},
\qquad \;\;\varepsilon_{\mu\nu} \varepsilon^{\mu\lambda} = 
- \;\delta_{\nu}^{\lambda}.
\end{array} \eqno (3.1)
$$
It turns out that the total 2D Lagrangian density (2.1)
$$
\begin{array}{lcl}
{\cal L}_{b} =
\frac{1}{2} E^2 - \frac{1}{2} (\partial \cdot A)^2 - i \partial_{\mu}
\bar C \partial^{\mu} C 
\end{array} \eqno (3.2)
$$  
remains invariant under the
combination of the above two transformations because the ghost term remains
invariant on its own and the kinetic energy term and gauge-fixing term
exchange with each-other: 
$$
\begin{array}{lcl}
\frac{1}{2} \; E^2 = \frac{1}{2} ( \partial_{0} A_{1}
- \partial_{1} A_{0} )^2\;\;\; \rightleftharpoons
- \frac{1}{2} (\partial \cdot A)^2 = \frac{1}{2}
(\partial_{0} A_{0} - \partial_{1} A_{1})^2.
\end{array} \eqno (3.3)
$$
Thus, in addition to the gauge BRST symmetry (2.2), we have an
on-shell ($ \Box \bar C = 0 $) nilpotent 
($ \delta_{d}^2 = 0 $) dual BRST
symmetry $\delta_{d}$ for the Lagrangian density (3.2)
$$
\begin{array}{lcl}
\delta_{d} A_{\mu} &=& - \eta\; \varepsilon_{\mu\nu}\; \partial^{\nu} \bar C,
\qquad\;\;\;\; 
\delta_{d} C = - i \eta E \nonumber\\
\delta_{d} E &=& \eta \Box \bar C, \quad 
\delta_{d} \bar C = 0, \quad 
\delta_{d} (\partial \cdot A) = 0 
\end{array}\eqno(3.4)
$$
which can be derived from (2.2) by the substitutions : 
$ C \rightarrow  + i \bar C, \partial_{\mu}
\rightarrow  + i \varepsilon_{\mu\nu} \partial^{\nu} $ 
\footnote{ Here and in what follows, we shall take only the ($+$) sign
in the transformations: $ C \rightarrow  \pm i \bar C, \bar C
\rightarrow  \pm i C, \partial_{\mu} \rightarrow  \pm i 
\varepsilon_{\mu\nu} \partial^{\nu} $. However, analogous
statements will be valid if we take ($-$) sign.}.  
We christen this symmetry as dual BRST 
because, in contrast to $\delta_{B}$ transformations where the electric field 
$E$ is invariant, in the case of $\delta_{D}$, it is the gauge-fixing
term $(\partial \cdot A)$ that remains invariant \footnote{ As per our 
definition in the introduction, the gauge-fixing term $\delta A = (\partial
\cdot A) $ with $ \delta = \pm {*} d {*} $ is the dual of the two-form 
$ F = d A$ which is the electric field $E$ here in 2D.}. Thus, we shall call
the duality transformations for the Lagrangian density (3.2) as the ones
where : $ C \rightarrow \pm\;i \bar C, \bar C \rightarrow \pm\;i C,
\partial_{\mu} \rightarrow \pm\;i \varepsilon_{\mu\nu} \partial^{\nu} $.
Introducing an auxiliary field $ {\cal B} $, the analogue of the
Lagrangian density (2.4) can be written as

$$
\begin{array}{lcl}
{\cal L}_{\cal B} = {\cal B}\; E - \frac{1} {2}\;{\cal B}^2 +
B \; (\partial \cdot A) + \frac{1}{2} B^2 - i \partial_{\mu} \bar C
\partial^{\mu} C
\end{array}\eqno(3.5)
$$
which respects the following off-shell nilpotent ($\delta_{D}^2 = 0$)
dual BRST symmetry

$$
\begin{array}{lcl}
\delta_{D} A_{\mu} &=& - \eta \varepsilon_{\mu\nu} \partial^{\nu} \bar C,
\quad \delta_{D} \bar C = 0,
\quad  \delta_{D} C = - i \eta {\cal B}, 
\quad \delta_{D} {\cal B} = 0 \nonumber\\
\delta_{D} E &=& \eta \Box \bar C, \;\;\;\;\;\qquad\;\;\;\;\;\;
\delta_{D} (\partial \cdot A) = 0,\;\;\; \qquad \;\;\;\;\delta_{D} B = 0.
\end{array}\eqno(3.6)
$$
This off-shell nilpotent dual BRST transformations
can be obtained from the transformations (2.5) 
(with the inclusion of $\delta_{B} {\cal B} = 0$)
by the substitution :
$ C \rightarrow + i \bar C, \partial_{\mu} \rightarrow + i
\varepsilon_{\mu\nu} \partial^{\nu}, B \rightarrow - i {\cal B},
 {\cal B} \rightarrow - i B$.
It can be checked that the off-shell nilpotent BRST and
dual BRST transformations (2.5) and (3.6) are connected with 
each-other by 

$$
\begin{array}{lcl}
C \rightarrow i \bar C, \quad  E \rightarrow i (\partial \cdot A),
\quad B \rightarrow - i\; {\cal B}  \nonumber\\
\bar C \rightarrow i C, \quad (\partial \cdot A)
\rightarrow i E, \quad {\cal B} \rightarrow - i\; B
\end{array}\eqno(3.7)
$$
which is a manifestation of the fact that the Lagrangian density (3.5) goes
to itself under the above substitutions. Thus, for the Lagrangian
density (3.5), the duality transformations are: $ C \rightarrow \pm\;i \bar C,
\bar C \rightarrow \pm\; i C, \partial_{\mu} \rightarrow \pm\;i
\varepsilon_{\mu\nu} \partial^{\nu}, B \rightarrow \mp\; i {\cal B},
{\cal B} \rightarrow \mp\; i B $
\footnote{Note that we have taken the upper sign of these transformations
in equation (3.7). However, the above statements are valid for the
lower sign as well.}. These continuous symmetries $ \delta_{(d,D)} $ lead
to the derivation of the following conserved and nilpotent 
($ Q_{(d,D)}^2 = 0$) dual BRST charge due to Noether theorem:
$$
\begin{array}{lcl}
Q_{(d,D)} = {\displaystyle \int} dx\;\bigl [ E \dot {\bar C} 
- \dot E \bar C \bigr ]
\equiv {\displaystyle \int } dx\;\bigl [ {\cal B} \;\dot {\bar C}
- \dot {\cal B}\; \bar C \bigr ] 
\end{array}\eqno(3.8)
$$
which generates (3.4) and (3.6) ({\it i.e.}, $\delta_{r} \phi = - i \eta
[ \phi, Q_{r} ]_{\pm}, r = d, D $ and $ (+)- $ stands for (anti)commutator 
corresponding to (fermionic)bosonic $\phi$). 
Due to discrete symmetry invariance of the ghost action
under $ C \rightarrow i \bar C, \bar C \rightarrow i C $, we have the
existence of a conserved and nilpotent anti-dual BRST charge 
$ Q_{(Ad, AD)}$ which can be derived
from (3.8) by these substitutions ({\it i.e.}, $ C \rightarrow i \bar C$).

It is obvious that $Q_{B}$ and $Q_{D}$ are the 
fermionic symmetry generators ($ Q_{B}^2 = 0, Q_{D}^2 = 0$)
for the Lagrangian density (3.5). Thus, the anticommutator of the two 
($ \{ Q_{B}, Q_{D} \}$) will also
be a symmetry generator. The corresponding
 bosonic 
transformation $\delta_{W} = \{ \delta_{B}, \delta_{D} \}$ 
with the infinitesimal bosonic transformation parameter $\kappa$
(= - i $\eta \;\eta^{\prime}$)

$$
\begin{array}{lcl}
\delta_{W} A_{\mu} &=& \kappa (\partial_{\mu} {\cal B}
+ \varepsilon_{\mu\nu} \partial^{\nu} B), \;\;\qquad\;\;
\delta_{W} B = 0, \;\;\qquad\;\;\; \delta_{W} {\cal B} = 0 \nonumber\\
\delta_{W} C &=& 0, \quad \delta_{W} \bar C = 0, \quad
\delta_{W} (\partial \cdot A) = \kappa \Box {\cal B}, \quad
\delta_{W} E = - \kappa \Box B
\end{array} \eqno (3.9)
$$
is the symmetry of the above Lagrangian density (3.5) because
$ \delta_{W} {\cal L}_{\cal B} = \kappa (\partial_{\mu} [\; B \partial^{\mu}
{\cal B} - {\cal B} \partial^{\mu} B \;] )$. Here $\eta$ and $\eta^{\prime}$
are the infinitesimal fermionic transformation parameters 
corresponding to $\delta_{B}$ and
$\delta_{D}$ respectively. The generator of the above symmetry
transformation (and the analogue of the Laplacian operator) is a 
conserved charge ($W$) given by:

$$
\begin{array}{lcl}
W = {\displaystyle \int} \; dx\; 
\bigl [\; {\cal B}\; \dot B - B\; \dot {\cal B}\; \bigr ].
\end{array}\eqno(3.10)
$$
This conserved quantity can be directly calculated from the anticommutator
$ \{ Q_{B}, Q_{D} \}$ if we exploit the canonical (anti)commutators :
$ \{ C(x,t), \dot {\bar C}(y,t) \} = \delta (x - y), \{ \bar C(x,t),
\dot C(y, t) \} = - \delta (x - y), [ A_{0} (x,t), B(y,t) ] = i \delta
(x - y), [ A_{1} (x,t), {\cal B} (y,t) ] = i \delta (x - y)$ and the rest
of the (anti)commutators are zero. Here $\delta (x - y)$ is the Dirac
delta function.\\

\noindent
{\bf 4 Extended BRST algebra}\\

\noindent
The set of all the conserved charges are the (anti)BRST, (anti)dual BRST,
ghost and the $W$ operator. Together, these charges for the 2D 
free $U(1)$ gauge theory are
$$
\begin{array}{lcl}
Q_{B} &=& {\displaystyle \int dx }\; 
\Bigl [\;B\;\dot C
-  \dot B \; C \;\Bigr ], \;\;\;\qquad \;\;\;
Q_{AB} = i {\displaystyle \int  dx} \;
\Bigl [\; B \;\dot {\bar C}
- \; \dot B\; \bar C \;\Bigr ] \nonumber\\
Q_{D} &=& {\displaystyle \int dx} \;\Bigl 
[\;{\cal B} \;\dot {\bar C} - \dot {\cal B}\; \bar C \;\Bigr ],
\;\;\qquad\;\;\;\;\; Q_{AD} = i 
{\displaystyle \int dx} \; 
\Bigl [\; {\cal B} \;\dot C - \dot {\cal B}\; C\; \Bigr]
\nonumber\\
W &=& {\displaystyle \int dx} \;
\Bigl [\;  {\cal B}\; \dot B - B\; \dot {\cal B}
\; \Bigr ],\;\;\; \qquad \;\;\;
Q_{g} = - i {\displaystyle \int dx} \;
\Bigl [\; C\; \dot {\bar C} + \bar C\; \dot C \;\Bigr ]. 
\end{array}\eqno(4.1)
$$
If we exploit the covariant canonical (anti)commutators,
these conserved charges obey the following extended BRST algebra
$$
\begin{array}{lcl}
&& [ W, Q_{k}] = 0, k = B, D, AB, AD, g \quad \nonumber\\
&&Q_{B}^2\; = \;Q_{AB}^2\; = \;Q_{D}^2\; = \;Q_{AD}^2\; = 0 \nonumber\\
&& \{ Q_{B}, Q_{D} \}\; = \;\{ Q_{AB}, Q_{AD} \}\; =\; W \nonumber\\
&& i [ Q_{g}, Q_{B} ] = + Q_{B}, \;\;\;\quad\;\;
i [ Q_{g}, Q_{AB} ] = - Q_{AB} \nonumber\\
&& i [ Q_{g}, Q_{D} ] = - Q_{D}, \qquad \;
i [ Q_{g}, Q_{AD} ] = + Q_{AD}
\end{array} \eqno(4.2)
$$
and all the rest of the (anti)commutators turn out to be zero.
A few remarks are in order. First of all, we see that the operator
$W$ is the Casimir operator for the whole algebra and its ghost
number is zero. The ghost number of $Q_{B}$ and $Q_{AD}$ is $+1$ and that 
of $Q_{D}$ and $Q_{AB}$ is $-1$. 
Now given a state $ |\psi>$ in the quantum Hilbert space with the ghost 
number $n$ ({\it i.e.}, $ i Q_{g} | \psi> = n |\psi >$), it is 
straightforward, due
to the above commutation relations, to check that the 
following relations are
satisfied:
$$ 
\begin{array}{lcl}
i Q_{g}\; Q_{B} |\psi> &=& (n + 1)\; Q_{B} |\psi> \nonumber\\
i Q_{g}\; Q_{D} |\psi> &=& (n - 1)\; Q_{D} |\psi> \nonumber\\
i Q_{g}\; W \;    |\psi> &=& n\; W \;|\psi>
\end{array}\eqno(4.3)  
$$ 
which demonstrate that, whereas $W$ keeps the ghost number of a state intact
and unaltered, the operator $Q_{B}$ increases the ghost number
by one and $Q_{D}$ reduces this number by one. This property is similar
to the operation of a Laplacian, an exterior derivative and a dual exterior
derivative on a $n$-form defined on a compact manifold. Thus,
we see that the degree of the differential form is analogous to the ghost
number in the Hilbert space, the differential form itself is analogous to the
quantum state in the Hilbert space, a compact manifold has an analogy with
the quantum Hilbert space and $d$, $\delta$ and $ \Delta = d \delta
+ \delta d $ are $Q_{B}$ , $Q_{D}$ and $W$ respectively. It is a notable
point that $d$ and $\delta$  can also be identified with
$Q_{AB}$ and  $Q_{AD}$ in the BRST formalism.\\

\noindent
{\bf 5 Hodge decomposition theorem and topological invariants}\\

\noindent
It is obvious from the algebra (4.2) and the consideration of 
the ghost number
of states ($Q_{B} |\psi>, Q_{D} |\psi> $ and $ W |\psi> $) in (4.3) 
that one can
now implement the Hodge decomposition theorem in the language of BRST and
dual-BRST charges (see, {\it e.g.}, [3], [7] ,[8])
$$
\begin{array}{lcl}
|\psi >_{n} = |\omega >_{n} +\; Q_{B} |\; \theta >_{n-1} +\; Q_{D}\; 
| \chi >_{n+1}
\end{array}\eqno(5.1)
$$
by which, any state $|\psi >_{n}$ in the quantum Hilbert space with 
ghost number $n$ can be decomposed into a harmonic state $|\omega >_{n}$,
a BRST exact state $ Q_{B} |\theta >_{n-1} $ and a dual-BRST exact state
$ Q_{D} | \chi >_{n +1} $. To refine the BRST cohomology, however, we have
to choose a representative state as the physical state from the
total states of the quantum Hilbert
space. Let us pick out here the physical state  
as the harmonic
state ($ | phys > = |\omega > $) from the Hodge decomposed state (5.1).
The number of such harmonic states is
finite for a given physical theory as it represents the number of solutions
to the Laplace equation (see, {\it e.g.}, Ref. [2]).
By definition, such a state would satisfy the following conditions:
$$
\begin{array}{lcl}
W |phys> = 0, \qquad Q_{B}\; | phys > = 0, \qquad Q_{D}\; | phys > = 0.
\end{array} \eqno(5.2)
$$
Due to the simple form of the equations of motion ($ \Box A_{\mu} = 0,
\Box C = 0, \Box \bar C = 0$) for the basic fields in the theory, it is very
convenient to express them in the normal modes [21]
$$
\begin{array}{lcl}
A_{\mu} (x, t) &=& {\displaystyle \int
\frac{ d k}{ (2 \pi)^{1/2} (2 k^0)^{1/2}}} \;
\Bigl [\; a_{\mu} (k) e^{- i k \cdot x} + a_{\mu}^\dagger(k) e^{i k\cdot x}\;
\Bigr ] \nonumber\\
C (x,t) &=& {\displaystyle \int \frac{dk}{ (2 \pi)^{1/2} (2 k^0)^{1/2}} }\;
\Bigl [\; c(k)\; e^{- i k \cdot x} \;+\; c^{\dagger}(k)\; e^{i k \cdot x} \;
\Bigr ] \nonumber\\
\bar C (x,t) &=& {\displaystyle \int \frac{d k} { (2 \pi)^{1/2} 
(2 k^0)^{1/2}}}\;
\Bigl [\; b(k)\; e^{- i k \cdot x} \;
+\; b^{\dagger}(k)\; e^{i k \cdot x} \;\Bigr ]
\end{array} \eqno(5.3)
$$
where $k_{\mu}$ are the 2D momenta with the components 
$( k_{0}, k_{1} = k)$. The
on-shell nilpotent symmetry transformations (2.2) and (3.4), 
that are generated
by the charges $Q_{b}$ and $Q_{d}$, can now be exploited to yield (see, 
{\it e.g.}, [21, 22] for details):
$$
\begin{array}{lcl}
&& [ Q_{b}, a_{\mu}^{\dagger}(k) ] = - k_{\mu}\; c^{\dagger}(k), \qquad
\;\;\;\;\; [ Q_{d}, a_{\mu}^{\dagger} (k) ] = \varepsilon_{\mu\nu} k^\nu
b^{\dagger} (k) \nonumber\\
&& [ Q_{b}, a_{\mu} (k) ] = k_{\mu} c(k), \;\;\;\;\qquad\;\;\;\;\;\; 
[ Q_{d}, a_{\mu} (k) ] = - \varepsilon_{\mu\nu} k^\nu b(k) \nonumber\\
&& \{ Q_{b}, c^{\dagger} (k) \} = 0, \;\;\; \;\;\;\;\;
\qquad\;\;\;\;\;\;\;\;\;
\{ Q_{d}, c^{\dagger} (k) \} = i \varepsilon^{\mu\nu} k_{\mu} 
a_{\nu}^{\dagger} \nonumber\\
&& \{ Q_{b}, c(k) \} = 0, \;\;\;\; \;\;\;\;\;
\qquad\;\;\;\;\;\;\;\;\; \{ Q_{d}, c(k) \}
= - i \varepsilon^{\mu\nu} k_{\mu} a_{\nu} \nonumber\\
&& \{ Q_{b}, b^{\dagger} (k) \} = - i k^{\mu} a_{\mu}^{\dagger},
\;\;\;\;\qquad\;\;\;\; \{Q_{d}, b^{\dagger} (k) \} = 0 \nonumber\\
&& \{ Q_{b}, b(k) \} = + i k^\mu a_{\mu}, \;\; \;\;\;\;\qquad\;\;\;\;
\{ Q_{d}, b(k) \} = 0. 
\end{array} \eqno(5.4)
$$
Similarly, the Casimir operator $W$ generates the following commutation
relations: 
$$
\begin{array}{lcl}
&& [ W, a_{\mu}^{\dagger} (k) ] =\; i k^2 
\varepsilon_{\mu\nu} (a^\nu)^{\dagger},
\qquad [ W, a_{\mu} (k) ] = - i k^2 \varepsilon_{\mu\nu} a^\nu \nonumber\\
&& [ W, c(k) ]\; = \;[ W, c^{\dagger} (k) ] \;=\; [ W, b(k) ]
= [ W, b^{\dagger} (k) ]\; =\; 0.
\end{array} \eqno(5.5)
$$
We are now in a position to define the physical vacuum $| vac> $ as
$$
\begin{array}{lcl}
Q_{b} |vac > &=& \;Q_{d} |vac > \;= \;W |vac >\; =\; 0 \nonumber\\
a_{\mu} |vac > &=& c(k) |vac > = \;b(k) |vac > = 0.
\end{array} \eqno(5.6)
$$
A single photon state $ | e(k), vac > $ with polarization vector $e_{\mu}$ can
be created from the physical vacuum by the application of a creation operator
$ e^{\mu} a_{\mu}^{\dagger} |vac > \equiv
| e(k), vac > $. The physicality criteria: $ Q_{b} |e (k), vac > = -
(k \cdot e) c^{\dagger} (k) | vac> = 0,\; Q_{d} | e(k), vac > = 
\varepsilon_{\mu\nu} e^\mu k^\nu b^{\dagger} (k) |vac > = 0 $ lead to the
transversality $(k \cdot e = 0)$ of the photon and the  
condition $ \varepsilon_{\mu\nu}
e^\mu k^\nu = 0 $ between $e_{\mu}$ and $k_{\mu}$.
Together, these conditions (due to the presence of extended symmetries) 
remove both the physical degrees of freedom of the 2D photon
and imply the masslessness condition $ k ^2 = 0$ 
(see, {\it e.g.}, Ref. [22] for more discussions).

The operation of
the $W$ operator on a single photon state ($i.e.$,
$ W |e(k), vac > = - i k^2 \varepsilon_{\mu\nu} e^\mu (a^\nu)^{\dagger}
|vac> = 0 $)
implies the on-shell condition ($ \Box A_{\mu} = 0 \rightarrow k^2 = 0 $) 
as well as the masslessness condition ($k^2 = 0$)
for the photon. The other relations: $ k \cdot e = 0, \varepsilon_{\mu\nu}
e^\mu k^\nu = 0,$ emerging from the operation of $Q_{b}$ and $Q_{d}$ on
a single photon state, are {\it unique} solutions to $k^2 = 0$. Thus, in
a subtle way, $W |phys> = 0$ does imply the validity of $Q_{b} |phys> = 0$
and $Q_{d} |phys> = 0$.  If basic symmetries are the central 
guiding principle, the operation of the $W$
operator on a single physical photon state in 2D is superfluous 
(in some sense) because the symmetry
corresponding to $W$ can be derived from the symmetries generated by
$Q_{(b,B})$ and $Q_{(d,D)}$. This fact is encoded in the 
expression for the operator
$W$ ({\it cf.} (4.1)) which can be re-expressed as
$$
\begin{array}{lcl}
W = {\displaystyle \int \; dx }\;
\frac{d} {dx} \;\bigl [ \; \frac{1}{2} \; {\cal B}^2 
- \;\frac{1}{2}\; B^2 \; \bigr ]
\rightarrow \; 0  \;\;\;\; \mbox{as} \;\;\;\;
x \rightarrow \pm \;\infty
\end{array}\eqno (5.7)
$$
due to the equation of motion $ \partial_{\mu} {\cal B}
+ \varepsilon_{\mu\nu} \partial^{\nu} B = 0 $. One can not think of
the off-shell validity of the expression for $W$ in (4.1) because of
the considerations of BRST cohomology. The presence of the two nilpotent 
symmetries corresponding to $Q_{(b,B)}$ and $Q_{(d,D)}$ and the 
requirement that:
$ Q_{(b,B)} |phys> = 0, Q_{(d,D)} |phys > = 0 $, 
forces the physical 2D photon
to always satisfy the on-shell ($ \Box A_{\mu} = 0 $) as well as the
mass-shell ($ k^2 = 0 $) condition. Thus, there is no escape from
the condition $W \rightarrow 0$ for a topological field theory where
all the physical degrees of freedom are gauged away by symmetries alone.
The topological nature of the theory is reflected by the presence of
the topological invariants on the 2D manifold. The two sets of these 
invariants, w.r.t. both the conserved ($ \dot Q_{B} = 0, \dot Q_{D} = 0$)
and off-shell nilpotent ($Q_{B}^2 = 0, Q_{D}^2 = 0$)
charges $Q_{B}$ and $Q_{D}$, are

$$
\begin{array}{lcl}
I_{k} [C_{k}] = {\displaystyle \oint}_{C_{k}} \; V_{k}, \qquad
J_{k} [C_{k}] = {\displaystyle \oint}_{C_{k}} \; W_{k} \qquad
(k = 0, 1, 2)
\end{array}\eqno (5.8)
$$
where $C_{k}$ are the k-dimensional homology cycles in the 2D manifold
and k-form $V_{k}$ and $W_{k}$ for the 2D free $U(1)$ gauge theory 
are juxtaposed as:
$$
\begin{array}{lcl}
V_{0} &=& B\;C, \;\;\;\;\;\;\;\;\;\;\;\;\;\;\qquad \;\;\;\;\;\;\;\;\;\;\;\; 
\;\;\;\;\;\;\;\;\;W_{0} = {\cal B}\; \bar C \nonumber\\
V_{1} &=& \bigl [\; B A_{\mu} + i C \partial_{\mu} \bar C\;
\bigr ]\; dx^{\mu},\;\; 
\qquad \;\;\;\;\;\;\;
W_{1} = \bigl [\; \bar C \varepsilon_{\mu\rho} \partial^{\rho} C
- i\; {\cal B} A_{\mu} \;\bigr ]\; dx^{\mu} \nonumber\\
V_{2} &=& i\bigl [  A_{\mu} \partial_{\nu} \bar C - \frac{1}{2} \bar C\;
F_{\mu\nu} \bigr ]\; dx^{\mu} \wedge dx^{\nu}, \quad
W_{2} = i \bigl [ \varepsilon_{\mu\rho} \partial^{\rho} C\; A_{\nu}
 + \frac{C}{2}\;\varepsilon_{\mu\nu} (\partial \cdot A) \bigr ]
\;dx^{\mu} \wedge dx^{\nu}.
\end{array}\eqno (5.9)
$$
It can be seen that $V_{0}$ and $W_{0}$ are BRST 
($\delta_{B} V_{0} = 0$) and co-BRST invariant ($\delta_{D} W_{0} = 0$)
and $V_{2}$ and $W_{2}$ are closed ($ d V_{2} = 0$) 
and co-closed ($ \delta W_{2} = 0$) respectively. Using the canonical
(anti)commutation relations with $i Q_{g}$, it can be checked that
the ghost numbers for $ (V_{0}, V_{1}, V_{2}) $  are ($ +1, 0, -1$)
and that of $(W_{0}, W_{1}, W_{2})$ are $(-1, 0, +1)$ respectively.
This fact can be succinctly expressed (for $ k = 0, 1, 2$) as:
$$
\begin{array}{lcl}
&& i [ Q_{g}, V_{k} ]\; = (-1)^{1-k}\; (k - 1)\; V_{k} \nonumber\\
&& i [ Q_{g}, W_{k} ] = (-1)^{1-k}\; (1 - k)\; W_{k}.
\end{array}\eqno (5.10)
$$
These invariants (for $ k = 1, 2 $) obey the following relations
(see, {\it e.g.}, [23], [24], [18])
$$
\begin{array}{lcl}
\delta_{B} V_{k} &=& \eta\; d\; V_{k-1}, \;\;\qquad\;\;\;
d = dx^{\mu} \;\partial_{\mu} \nonumber\\
\delta_{D} W_{k} &=&  \eta\; \delta \; W_{k-1}, \;\;\qquad\;\;
\delta = i\; dx^{\mu}\; \varepsilon_{\mu\nu}\;
\partial^{\nu}
\end{array}\eqno (5.11)
$$
where $d$ and $\delta$ are the exterior and dual-exterior derivatives on
the 2D compact manifold. Both these sets of topological invariants are
related to each-other by the duality transformations (3.7) as
$ I_{k} \rightarrow J_{k} $ under the sustitutions: $ B \rightarrow
-i\; {\cal B}, C \rightarrow i \bar C, \partial_{\mu} \rightarrow
 i\; \varepsilon_{\mu\nu} \partial^{\nu} $.

Using the on-shell nilpotent BRST- and dual-BRST transformations (2.2)
and (3.4), it will be interesting to verify that, modulo some total 
derivatives,
the  Lagrangian density (3.2) can be
written as the sum of BRST- and co-BRST invariant parts:
$$
\begin{array}{lcl}
\eta\; {\cal L}_{b} = \frac{1}{2} \;
\delta_{d} \bigl [\; i E C \;\bigr ] - \frac{1}{2} \;
\delta_{b} \bigl [\; i (\partial \cdot A) \bar C \;\bigr ].
\end{array} \eqno(5.12)
$$
The invariance of this Lagrangian density under BRST and dual BRST 
transformations is easy to see because $ \delta_{b}^2 = 0, \delta_{d}^2 = 0$
and $ \{ \delta_{d}, \delta_{b} \} \rightarrow 0 $ as the Laplacian operator
goes to zero ($ W \rightarrow 0$) for the validity of the equations of motion.
Furthermore, the expressions in the square brackets in (5.12)
are BRST invariant
({\it i.e.}, $ \delta_{b} [i E C] = 0$) and co-BRST invariant 
({\it i.e.}, $\delta_{d}
[ i (\partial \cdot A) \bar C] = 0$). 
Using the fact that $Q_{r}\; (r = b,d)$ is the generator of transformation
$ \delta_{r} \phi = - i\;\eta\;[ \phi, Q_{r}]_{\pm},
$ where $ (+)- $ stands for the (anti)commutator corresponding to
$\phi$ being (fermionic)bosonic in nature, it can be seen that (5.12) can be
written as:
$ {\cal L}_{b} = \{ Q_{d}, S_{1} \} + \{ Q_{b}, S_{2} \} $ for $ S_{1}
= \frac{1 }{2}\; E C, S_{2} = - \frac{1} {2}\; (\partial \cdot A) \bar C $.
This shows that the free $U(1)$ topological 
gauge field theory is similar {\it in form} as the
Witten type theories [24] but completely different in outlook from the
Schwarz type theories [25]. To be very precise, the free $U(1)$
topological gauge field theory is somewhat different
from Ref. [24] too.  This is mainly because of the fact that, 
in our discussions,
there are two conserved and nilpotent charges w.r.t. which the topological
invariants are defined whereas in Ref. [24] there exists only a single BRST
charge which is obtained due to the presence of topological shift- and
local gauge symmetries. In our discussions, there is no shift symmetry
at all. Thus, from {\it symmetry point of view}, the 2D free $U(1)$ gauge
theory is more like Schwarz type theories. It can be seen, however, that
the {\it symmetric} energy-momentum tensor ($T_{\mu\nu}$) for the Lagrangian 
density (3.2) (or (5.12))
$$
\begin{array}{lcl}
T_{\mu\nu} &=& 
- \frac{1}{2} \bigl [\; \varepsilon_{\mu\rho} E + \eta_{\mu\rho}
(\partial \cdot A) \;\bigr ] \partial_{\nu} A^{\rho}
- \frac{1}{2} \bigl [\; \varepsilon_{\nu\rho} E + \eta_{\nu\rho}
(\partial \cdot A) \;\bigr ] \partial_{\mu} A^{\rho}\nonumber\\
&-& i \partial_{\mu} \bar C \partial_{\nu} C - i \partial_{\nu} \bar C
\partial_{\mu} C - \eta_{\mu\nu}\; {\cal L}_{b},
\end{array}\eqno(5.13)
$$
has the {\it same form} as the Witten- and Schwarz type of 
topological field theories because it can be re-expressed as:
$$
\begin{array}{lcl}
T_{\mu\nu} = \{ Q_{b}, V^{(1)}_{\mu\nu} \} + \{ Q_{d}, V^{(2)}_{\mu\nu} \},
\end{array}\eqno(5.14)
$$
where the exact expression for $V's$, in terms of the local fields, are
$$
\begin{array}{lcl}
V^{(1)}_{\mu\nu} &=& \frac{1}{2} \bigl [ \;\partial_{\mu} \bar C \;A_{\nu} 
+ \partial_{\nu} \bar C \;A_{\mu} + \eta_{\mu\nu} (\partial \cdot A) 
\bar C  \;\bigr ], \nonumber\\
V^{(2)}_{\mu\nu} &=& \frac{1}{2} \bigl [\; \partial_{\mu}  C 
\varepsilon_{\nu\rho} A^{\rho} 
+ \partial_{\nu}  C \varepsilon_{\mu\rho} A^{\rho}
- \eta_{\mu\nu} \; E C  \;\bigr ].
\end{array}\eqno(5.15)
$$ 
It 
can be checked that the partition functions as well as the expectation values
of the BRST invariants, co-BRST invariants and the topological invariants
are metric  independent
\footnote{ We have taken here only the flat Minkowski metric. However,
our arguments are valid even if we take into account a nontrivial metric.
The metric independence of the measure has been shown in Ref. [23].}.
The key point to show this fact in the framework of BRST cohomology is the
requirement that $Q_{b} | phys > = 0$ and $Q_{d} | phys > = 0 $
(see, {\it e.g.}, Ref. [18] for details) and the metric independence of the
path integral measure (see, {\it e.g.}, Ref. [23]).\\

\noindent
{\bf  6 Conclusions}\\

\noindent
It is obvious  that the usual nilpotent BRST transformations
correspond to a symmetry in which the two-form $ F = d A $
($e.g.$, electric field $E$ in 2 D) of the $U(1)$ gauge theory remains 
invariant.
The nilpotent dual-BRST charge is the generator of a transformation in which
the gauge-fixing term $((\partial \cdot A)= \delta  A)$ remains invariant.
The anticommutator of these two transformations corresponds to a symmetry that 
is generated by the Casimir operator for the whole algebra. Under this 
conserved operator, it is the ghost term that remains invariant.
Basically, the presence of BRST- and dual BRST symmetries imply
the existence of two gauge symmetries:
$ e_{\mu}
\rightarrow  e_{\mu} + \alpha \; k_{\mu}, e_{\mu} \rightarrow
e_{\mu} + \beta\; \varepsilon_{\mu\nu} \; k^{\nu} $
(for $\alpha$ and $\beta$ being arbitrary constants)
in the theory. In the present work, these extended 
symmetries have been exploited together to gauge away the dynamical degrees of 
freedom of 2D photon so that this theory becomes topological. 
The form of the Lagrangian density (5.12), the appearance of
symmetric energy-momentum tensor (5.14) and the existence of BRST-
and co-BRST invariants in (5.9) confirm this (topological) 
nature of the theory. 
In fact, it is a new type of topological
field theory which captures together 
some of the salient features of both Witten-
and Schwarz type of theories. It is an interesting venture to
generalize these symmetries
to 2D free (having no interaction with matter fields) [26]
as well as interacting non-Abelian gauge theories. 
Furthermore, it will be nice to explore the physical
impact of these kind of symmetries in the context of physical
4D interacting gauge theories.
In fact, as a first preliminary step in this direction,
it has been 
shown in Ref. [19] that the dual-BRST transformation $ \delta_{D} A_{\mu} 
= - \eta \varepsilon_{\mu\nu} \partial^{\nu} \bar C $ 
on the Abelian gauge field corresponds to the 
chiral transformation on the Dirac fields for fermions in 2D
interacting $U(1)$ gauge theory. Thus, 
the ABJ anomalies appear in the theory 
for the proof of conservation laws at the quantum level. It
is, therefore, expected that the full strength of
BRST cohomology  and Hodge decomposition theorem  
might shed some light on the ABJ anomalies and
provide a clue to the well known result 
that in 2D, the ``anomalous'' gauge theory is consistent, unitary 
and amenable to particle interpretation [27,28]. The insights gained in 2D
might turn out to be useful for the generalization of Hodge decomposition
to physical 4D gauge theories.
These are some of the issues which are under investigation and the results
will be reported elsewhere.\\

\baselineskip = 12pt

\noindent {\bf References}
\begin{enumerate}
\item   Eguchi T, Gilkey P B  and  Hanson A J 1980 
           {\it Phys. Rep.}  {\bf 66} 213  
\item   Mukhi S and  Mukunda N 1990  {\it Introduction  to Topology,
           Differential Geometry and Group Theory for Physicists} 
           (Wiley Eastern
           Ltd.: New Delhi) 
\item   van Holten J W 1990 {\it Phys. Rev. Lett.} {\bf 64} 2863 
\item   Aratyn H 1990  {\it J. Math. Phys.}   {\bf 31} 1240 
\item   Dirac P A M 1964 {\it Lectures on Quantum Mechanics} (Yeshiva
           University Press: New York)
\item   Nishijima K 1986 in:   {\it Progress in Quantum Field Theory} eds.
            Ezawa H
           and  Kamefuchi S (North- Holland: Amsterdam) 
\item   Henneaux M and Teitelboim C 1992 
           {\it Quantization of Gauge Systems} 
           (Princeton University Press: Princeton)
\item   Nakanishi N and  Ojima I 1990 {\it Covariant Operator Formalism of
           Gauge Theories and Quantum Gravity} 
          (World Scientific: Singapore)
\item   Gitman D M and Tyutin I V 1990  {\it Quantization of Fields
           with Constraints}  (Springer-Verlag: Berlin)
\item   Sundermeyer K 1982  {\it Constrained Dynamics, Lecture Notes
           in Physics} (Springer-Verlag: Berlin)
\item   Batalin I A and Tyutin I V 1991 {\it Int. J. Mod. Phys.} 
           {\bf A6} 3255 
\item   Batalin I A,
           Lyakhovich S L and Tyutin I V 1992
           {\it Mod. Phys. Lett.} {\bf A7}
           1931; 1995 {\it Int. J. Mod. Phys.} {\bf A10} 1917
\item   McMullan D and Lavelle M 1993 {\it Phys. Rev. Lett.} {\bf 71} 
           3758; 1995
           {\it ibid.}  {\bf 75} 4151
\item   Rivelles V O 1995 {\it Phys. Rev. Lett.}  {\bf 75} 4150;
           1996 {\it Phys. Rev.} {\bf  D53} 3257
\item   Yang H S and Lee B -H 1996   {\it J. Math. Phys.}
           {\bf 37} 6106 
\item   Marnelius R 1997   {\it Nucl. Phys.} {\bf B494} 346
\item   Zhong T and Finkelstein D 1994 {\it Phys. Rev. Lett.}
           {\bf 73} 3055; 1995  {\it ibid.}  {\bf 75} 4152
\item   Birmingham D,  Blau M, Rakowski M and Thompson G 1991
           {\it Phys. Rep.}  {\bf 209} 129 
\item   Malik R P  {\it Dual BRST Symmetry in QED} :
            hep-th/ 9711056   
\item   Chryssomalakos C, de Azcarraga J A, Macfarlane A J
           and Perez Bueno J C  {\it Higher order BRST and anti-BRST
           operators and cohomology for compact Lie algebras} :
           hep-th/9810212
\item   Weinberg S 1996
           {\it The Quantum Theory of Fields: Modern Applications}, 
           {\bf V.2} (Cambridge University Press : Cambridge)
\item   Malik R P  {\it On the BRST cohomology in $U(1)$ gauge theory} :
            hep-th/ 9808040 (To appear in: 2000
            {\it Int. J. Mod. Phys. {\bf A}})
\item   Kaul R K and  Rajaraman R 1991 {\it Phys. Lett.}  {\bf B265}
           335; 1990 {\it ibid.} {\bf B249} 433 
\item   Witten E 1989 {\it Commun. Math. Phys.}  {\bf 121} 351
\item   Schwarz A S 1978 {\it Lett. Math. Phys.} {\bf 2} 217 
\item   Malik R P 1999 {\it Mod. Phys. Lett.} {\bf A14} 1937
\item   Jackiw R and  Rajaraman R 1985  {\it Phys. Rev. Lett.}  
           {\bf 54} 1219
\item   Malik R P 1988 {\it Phys. Lett.} {\bf B212} 445
\end{enumerate}
\end{document}